\documentclass[sigconf]{acmart}
\usepackage{enumitem}

\usepackage{amssymb}
\usepackage[ruled]{algorithm2e}
\usepackage{url}
\usepackage{multirow}
\usepackage{subfigure}
\usepackage{bm}
\usepackage{verbatim}
\usepackage{ragged2e}
\usepackage{caption}
\usepackage{subcaption}
\usepackage{graphicx}
\usepackage[normalem]{ulem}
\useunder{\uline}{\ul}{}
\usepackage{amsmath}
\usepackage{longtable}
\usepackage{adjustbox}
\usepackage{microtype}
\usepackage{setspace}
\usepackage{hyperref}
\usepackage{float}
\usepackage{balance}
\newcommand{\ie}{\emph{i.e., }}
\newcommand{\eg}{\emph{e.g., }}

\newcommand{\aka}

\AtBeginDocument{%
  \providecommand\BibTeX{{%
    \normalfont B\kern-0.5em{\scshape i\kern-0.25em b}\kern-0.8em\TeX}}}

\copyrightyear{2024}
\acmYear{2024}
\setcopyright{acmlicensed}
\acmConference[WSDM '24] {Proceedings of the 17th ACM International Conference on Web Search and Data Mining}{March 4--8, 2024}{Merida, Mexico.}
\acmBooktitle{Proceedings of the 17th ACM International Conference on Web Search and Data Mining (WSDM '24), March 4--8, 2024, Merida, Mexico}
\acmPrice{15.00}
\acmISBN{979-8-4007-0371-3/24/03}
\acmDOI{10.1145/3616855.3635816}

\begin{document}

\newcommand{\bym}[1]{\textcolor{black}{#1}}
\title{LabelCraft: Empowering Short Video Recommendations with Automated Label Crafting}

\author{Yimeng Bai*}
\orcid{0009-0008-8874-9409}
\affiliation{
  \institution{University of Science and Technology of China}
  \city{Hefei}
  \country{China}
}
\email{baiyimeng@mail.ustc.edu.cn}
\thanks{*Work done at Kuaishou.}

\author{Yang Zhang$^{\dag}$}
\orcid{0000-0002-7863-5183}
\affiliation{
  \institution{University of Science and Technology of China}
  \city{Hefei}
  \country{China}
}
\email{zy2015@mail.ustc.edu.cn}

\author{Jing Lu}
\orcid{0009-0000-0718-6766}
\affiliation{
  \institution{Kuaishou Technology}
  \city{Beijing}
  \country{China}
}
\email{lvjing06@kuaishou.com}

\author{Jianxin Chang}
\orcid{0000-0002-7886-9238}
\affiliation{
  \institution{Kuaishou Technology}
  \city{Beijing}
  \country{China}
}
\email{changjianxin@kuaishou.com}

\author{Xiaoxue Zang}
\orcid{0000-0002-5923-3429}
\affiliation{
  \institution{Kuaishou Technology}
  \city{Beijing}
  \country{China}
}
\email{zangxiaoxue@kuaishou.com}

\author{Yanan Niu}
\orcid{0000-0003-2083-518X}
\affiliation{
 \institution{Kuaishou Technology}
 \city{Beijing}
 \country{China}}
\email{niuyanan@kuaishou.com}

\author{Yang Song}
\orcid{0000-0002-1714-5527}
\affiliation{
  \institution{Kuaishou Technology}
  \city{Beijing}
  \country{China}
}
\email{yangsong@kuaishou.com}

\author{Fuli Feng$^{\dag}$}
\orcid{0000-0002-5828-9842}
\affiliation{%
  \institution{University of Science and Technology of China \& USTC Beijing Research Institute}
  \city{Hefei}
  \country{China}
}
\email{fulifeng93@gmail.com}
\thanks{$^{\dag}$Corresponding author.}

\def\authors{Yimeng Bai, Yang Zhang, Jing Lu, Jianxin Chang, Xiaoxue Zang, Yanan Niu, Yang Song, Fuli Feng}

\renewcommand{\shortauthors}{Yimeng Bai et al.}

\begin{abstract}

Short video recommendations often face limitations due to the quality of user feedback, which may not accurately depict user interests. To tackle this challenge, a new task has emerged: generating more dependable labels from original feedback. Existing label generation methods rely on manual rules, demanding substantial human effort and potentially misaligning with the desired objectives of the platform. To transcend these constraints, we introduce \textit{LabelCraft}, a novel automated label generation method explicitly optimizing pivotal operational metrics for platform success. By formulating label generation as a higher-level optimization problem above recommender model optimization, LabelCraft introduces a trainable labeling model for automatic label mechanism modeling. Through meta-learning techniques, LabelCraft effectively addresses the bi-level optimization hurdle posed by the recommender and labeling models, enabling the automatic acquisition of intricate label generation mechanisms. Extensive experiments on real-world datasets corroborate LabelCraft’s excellence across varied operational metrics, encompassing usage time, user engagement, and retention. Codes are available at \url{https://github.com/baiyimeng/LabelCraft}.

\end{abstract}

\ccsdesc[500]{Information systems~Recommender systems}

\keywords{Short Video Recommendation; Label Generation}



\maketitle

\section{Introduction}\label{intro}

Short video platforms including TikTok and YouTube Shorts are incredibly popular worldwide, making short video recommender systems vital for information filtering on the Web~\cite{MTIN,UCRAFB, DRM}. Within these platforms, users mostly offer implicit feedback (watch time) by scrolling and watching videos, occasionally adding explicit feedback like comments and likes~\cite{RLUR,ActorCriticRec}. The mixed feedback establishes the essential data foundation for building short video recommender systems~\cite{ActorCriticRec}. However, directly using the feedback as the label might lead to biased insights into user preferences due to its unreliability as an indicator~\cite{DVR,D2Q,DML,DCR}. For instance, watching 15 seconds of a 60-second video doesn't necessarily indicate a stronger preference than watching a 5-second video \textit{repeatedly} for 10 seconds, despite longer watch time. Therefore, the task of generating labels from raw feedback has emerged as a pivotal endeavor for building adept short video recommender systems~\cite{DML}.

Previous research has predominantly relied on rule-based strategies for label generation in the context of short video recommendations. Certain approaches focus on transforming user feedback, such as watch time, into new semantic labels that capture hidden signals of user preferences~\cite{DML, PC, PCR}. Other methods attempt to establish rules for refining original feedback, with a focus on eliminating bias in labels~\cite{DML, DVR, D2Q}. These manually generated labels, along with other feedback like comments, are then utilized in a multi-task learning framework to improve recommendations \cite{ActorCriticRec}. However, these rule-based methods require notable manual effort and may not consistently align with the operational metrics~\cite{ResAct} of the recommender system, such as user engagement. For example, the proposed unbiased watch time label in~\cite{D2Q} has been observed to reduce user engagement, evidenced by a reduction in "share"~\cite{D2Q}.

Considering the limitations of current approaches, we pursue automated label crafting and suggest reframing the label generation process as an optimization problem. Specifically, the label generation process should explicitly target optimizing the platform objectives, which are reflected by operational metrics. These objectives provide fundamental guidance for designing an effective recommender system \cite{ActorCriticRec}. As such, by aligning the label generation process with the objective optimization, we could expect to achieve a superior recommender system. Additionally, this innovative formulation allows for label generation to be performed through a learning-based approach, eliminating the need for labor-intensive manual creation of labeling rules. Consequently, this opens up possibilities for a more delicate mechanism of label generation with consideration of more influencing factors.

In this work, we propose \textit{LabelCraft}, a novel automated label generation approach that prioritizes explicit optimization aligned with the platform objectives. We use a trainable labeling model that considers all possible factors (\textit{e.g.}, all user feedback) to create the label, enabling comprehensive utilization of these influencing factors. To introduce the desired explicit optimization for the platform objectives, we formulate the learning process of the recommender and label models as a bi-level optimization problem. The first level involves fitting the generated labels to train the recommender, while the second level focuses on learning the labeling model by optimizing the acquired recommender model's performance with respect to the platform objectives. By employing meta-learning techniques~\cite{SML,MAML}, LabelCraft could effectively address the optimization challenge, leading to the acquisition of increasingly sophisticated mechanisms for label generation.

After reviewing previous work on short video recommendations~\cite{D2Q, DVR, DML, lin2023tree, wang2022surrogate, stamenkovic2022choosing, ActorCriticRec, liu2023generative}, we identify three main aspects (operational metrics) of industrial platform objectives for recommendations: user usage time, user engagement, and user retention. From the perspective of recommendation results, the two former aspects are reflected by the total count of the watch time \cite{D2Q, DVR, lin2023tree, DML} and positive explicit feedback \cite{ActorCriticRec, liu2023generative} for a top-$k$ recommended list, while user retention is highly affected by the recommendation list diversity~\cite{wang2022surrogate, stamenkovic2022choosing}. Therefore, we approximately represent the platform objectives using these list-wise quantities. During training, we utilize the soft top-$k$ technique \cite{xie2020differentiable} to compute them, enabling gradient updates, and we employ dynamic balancing schemes to prevent the skewed optimization of the different aspects. We demonstrate our method on \textit{Deep Interest Network} (DIN) \cite{DIN} and conduct experiments on two real-world datasets, validating its capability to consistently achieve favorable outcomes across all aspects of the platform objectives.

The main contributions of this work are summarized as follows:
\begin{itemize}[leftmargin=*]
    \item We emphasize the importance of label generation research for short video recommendations and formulate the label generation process as an optimization problem.  
    \item We introduce LabelCraft, an automated and effective label generation approach that explicitly optimizes the platform objectives utilizing learning techniques.
    \item We conduct extensive experiments on two real-world datasets. Extensive results demonstrate the effectiveness of our proposal.
\end{itemize}
\section{Task Formulation}

\paragraph{Label generation} Let $\mathcal{D}$ represent the historical data. Each sample in $\mathcal{D}$ is denoted as $(\bm{x},\bm{y}^{r})$, where $\bm{x}=[x_1,\dots,x_{d}] \in \mathcal{X}$ describes the features of the user-item pair and $\bm{y}^{r}=[y_1,\dots,y_{w}] \in \mathcal{Y}^{r}$ represents diverse raw user feedback on the video. For convenience in the following context, $x_d$ specifically signifies the video duration, $y_w$ is specifically indicative of the watch time feedback. Label generation is a process before recommendation learning, of which the core is to find a labeling model $g(\cdot)$ that maps raw feedback to a new label $y^{c}\in \mathcal{Y}^{c}$ with considering the influence of feature $\bm{x}$. In this work, we consider finding a labeling model in a learning manner, so we parameterize it with $\phi$, and re-denote it as $g_{\phi}$.   
Formally, we need to find a
\begin{equation} \label{eq:label_map}
     g_{\phi}: \mathcal{X} \times \mathcal{Y}^{r}  \longmapsto \mathcal{Y}^{c},
\end{equation}
and expect that the label generated by it is helpful for building a better recommender model regarding platform objectives such as user engagement. \bym{Without loss of generality, we assume these objectives are given by the platform and can be measured.}

\paragraph{Recommendation} 
The target is to learn a recommender model $f_{\theta}$ that uses $\bm{x}$ to predict the considered label by fitting historical data, where $\theta$ denotes its model parameters. 
In this work, we hope to solely fit the generated label, which means we need to learn a
 \begin{equation}\label{eq:reco_map}
    f_{\theta}: \mathcal{X} \longmapsto \mathcal{Y}^{c}, 
 \end{equation}
which could accurately predict $y^c$ 
for candidate items, so as to construct recommendations aligned with platform objectives.
\section{Methodology}

In this section, we first provide an overview of the proposed LabelCraft, followed by its detailed learning strategy. Lastly, we present our designs of the platform objectives. 
\begin{figure}[t]
    \centering
    \includegraphics[width=0.46\textwidth]{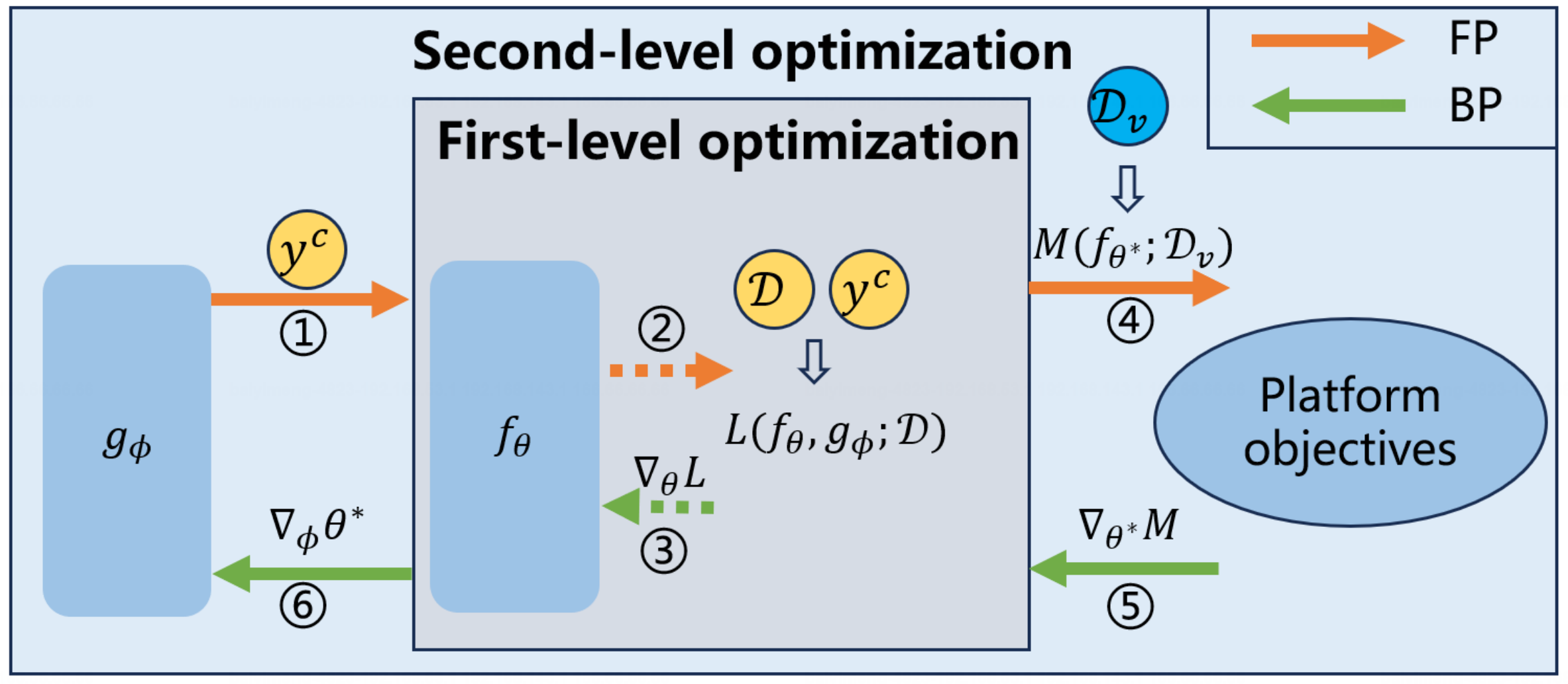}
    \caption{An overview of LabelCraft, which utilizes the platform objectives to guide the learning of the labeling model $g_{\phi}$ with the recommender model $f_{\theta}$ as the intermediary bridge.}
    \label{fig:framework}
\end{figure}

\subsection{Overview}
 
We strive to address the label generation task outlined in Equation~\eqref{eq:label_map} in an automated and efficient manner. As depicted by the equation, the primary hurdle stems from the absence of direct guidance for acquiring the mapping function due to the lack of ground truth for the target label.
To surmount this challenge, we hold the belief that the platform objectives serves as foundational guidance for devising an effective recommender system, thus naturally providing guidance for generating labels.

With this belief, we propose an automated label generation approach called LabelCraft, as illustrated in Figure~\ref{fig:framework}.  The core is formulating a bi-level optimization problem, where the first level optimizes the recommender model with the labels generated by the labeling model, while the second level optimizes the labeling model according to the performance of the learned recommender on the platform objectives. 
Then, by using meta-learning \cite{MAML} to solve the problem,  we successfully leverage the platform objectives to guide the learning of labeling model $g_\phi$, with the recommender model $f_\theta$ as an information (gradient) passing bridge between them. 
Next, we provide the details of the formulated optimization problem and the learning strategy to solve the problem.

\subsection{\textbf{Bi-level Optimization Problem}} To introduce explicit optimization towards the platform objectives for label generation, we formulate a bi-level optimization problem about $g_\phi$ and $f_\theta$ as follows:
\begin{subequations}\label{eq:overall-obj} \small
\begin{align}
 & \max_{\phi} \, \mathit{M}(f_{\theta^*}; \mathcal{D}_v) \label{eq:over-a} \\
 \textit{s.t.} \quad
& \theta^* = \mathop{\arg\min}_{\theta} \, L(f_{\theta}, g_{\phi};\mathcal{D}) \label{eq:over-b},
\end{align}
\end{subequations}
where $L(f_{\theta}, g_{\phi};\mathcal{D})$ denotes the recommendation loss of $f_{\theta}$ on the training data $\mathcal{D}$ by fitting the label generated by $g_{\phi}$; $\theta^{*}$ denotes the learned recommender model parameters by minimizing $L(f_{\theta}, g_{\phi};\mathcal{D})$ and is dependent on $\phi$; and $\mathit{M}(f_{\theta^{*}}; \mathcal{D}_v)$ denotes the recommendation performance of the learned model
$f_{\theta^*}$ in terms of platform objectives measured on a hold-out dataset $\mathcal{D}_{v}$. Formally, we have
\begin{equation}
\begin{split}
       L(f_{\theta}, g_{\phi};\mathcal{D}) = \frac{1}{|\mathcal{D}|}\sum_{(\bm{x},\bm{y}^{r}) \in \mathcal{D}} l(f_\theta(
\bm{x}
),y^{c}) + \lambda \|\theta\|^2 ,
 \quad y^c = g_{\phi}(\bm{x},\bm{y}^{r}),
\end{split}
\end{equation}
where $\lambda$ is a coefficient to adjust $L_2$ regularization for preventing overfitting, and
$l(\cdot)$ denotes a recommendation loss such as MSE~\cite{ding2022interpolative}. Regarding $\mathit{M}(f_{\theta^*}; \mathcal{D}_v)$ to measure the performance on the platform objectives, we measure it with some operational metrics, which will be explained later.

Obviously, Equation~\eqref{eq:over-b} aims at learning a recommender model that minimizes the loss to fitting the label generated by $g_\phi$, which constructs the first-level optimization. Equation~\eqref{eq:over-a} aims at finding a labeling model $g_\phi$, such that the recommender model fitting its generated labels could achieve the best recommendation performance measured by $M(\cdot)$. 
The bi-level optimization problem provides the mathematical formulation for the label generation problem. It successfully bridges the platform objectives and the labeling model using the recommender model $f_{\theta^*}$ learned on generated labels, facilitating the explicit optimization of the labeling model towards the platform objectives.

\begin{algorithm}[t]
    \caption{Training of LabelCraft}
    \label{alg:opt}
    \LinesNumbered
    \KwIn{Recommender model $f_\theta$, labeling model $g_\phi$, training dataset $\mathcal{D}$, hold-out dataset $\mathcal{D}_v$, recommender learning rate $\eta_1$ for $f_{\theta}$, and  learning rate $\eta_2$ for $g_{\phi}$.}
    Initialize $\phi$ and $\theta$ randomly\;
    \While{Stop condition is not reached}{
        {
        // Step 1 (update of $\phi$)\;
        Compute $\theta^{\prime}$ with Equation~\eqref{eq:step1}\;
        Update $\phi$ according to Equation ~\eqref{eq:step2}\;
        // Step 2 (update of $\theta$)\;
        Update $\theta$ according to Equation ~\eqref{eq:step3}\;
        }
    }
    return $f_\theta, g_\phi$
\end{algorithm}
\subsection{Learning Strategy}
The two levels of optimization in Equation~\eqref {eq:overall-obj} affect each other, forming a loop of nested optimization. Directly solving the total optimization problem is challenging.  To overcome this challenge, we develop a training method based on the meta-learning technique --- 
 MAML~\cite{MAML}, which considers alternatively updating the $\phi$ and $\theta$ in a loop.  For each interaction, we update $\phi$ and $\theta$ as follows:

 \begin{enumerate}[leftmargin=8pt]
    \item[-] \textbf{Update $\phi$}. To update the labeling model, finding the corresponding $\theta^{*}$ involves the full training of $\theta$, which is expensive. Instead of fully optimizing it, we just make a tentative update of the recommender model using the training data $\mathcal{D}$, following previous works~\cite{chen2021autodebias}. Then inspect the performance of the obtained recommender model about the platform objectives using the hold-out dataset $\mathcal{D}_{v}$. The gotten performance is used to update the labeling model. To be exact, there are two steps:
    \begin{itemize}[leftmargin=8pt]
        \item Step 1. Meta training. First, we keep the labeling model unchanged and fictively update the recommender model. In particular, we fix the parameter $\phi$ and compute the recommendation loss on the training dataset $\mathcal{D}$, \ie $L(f_{\theta}, g_{\phi}; \mathcal{D})$ in Equation~\eqref{eq:over-b}. Then we make a 
        step of gradient descent to obtain a tentative recommender model $f_{\theta^{\prime}}$. Formally, we have
        \begin{equation}\label{eq:step1}
            \theta^{\prime} =\theta-\eta_1 \nabla_\theta L(f_{\theta}, g_{\phi}; \mathcal{D}),
        \end{equation}
        where $\eta_{1}$ denotes the learning rate for the recommender model. 

        \item Step 2. Meta testing. After getting the tentative recommender model $f_{\theta^{\prime}}$, we evaluate its performance in terms of the platform objectives on the hold-out dataset $\mathcal{D}_{v}$, getting $\mathit{M}(f_{\theta^{\prime}}; \mathcal{D}_{v})$ similar to $\mathit{M}(f_{\theta^{*}}; \mathcal{D}_{v})$ in Equation~\eqref{eq:over-a}. Then we update $\phi$ to maximize $\mathit{M}(f_{\theta^{*}}; \mathcal{D}_{v})$, formally,
        \begin{equation}\label{eq:step2}
            \phi \leftarrow \phi + \eta_2 \nabla_\phi M(f_{\theta^{\prime}}; \mathcal{D}_v),
        \end{equation}
        where $\eta_2$ is the learning rate for the labeling model. The gradients here can be computed by using the back-propagation along the chain $M(f_{\theta^{\prime}}; \mathcal{D}_v) \rightarrow \theta^{\prime} \rightarrow \nabla_{\theta} L(f_{\theta}, g_{\phi}; \mathcal{D}) \rightarrow \phi$.

    \end{itemize}

    \item[-] \textbf{Update $\theta$}. After the update of $\phi$, we obtain a new labeling model. Consequently, we use it to generate new labels and optimize the recommendation loss to update $\theta$. This update is formulated as:
    \begin{equation}\label{eq:step3}
        \theta\leftarrow\theta-\eta_1 \nabla_\theta L(f_\theta, g_\phi; \mathcal{D}),
    \end{equation}
\end{enumerate}

The above two updates are iterated until convergence. 
Algorithm \ref{alg:opt} summarizes the detailed learning algorithm. 
In each interaction, it first updates $\phi$  (line 4-5) and then updates $\theta$ (line 7), and in the implementation, the updates are performed on a batch of data.  

\subsection{Platform Objectives}
In this subsection, we outline our approach for evaluating a recommender model's performance with respect to platform objectives. To address the expensive cost of the online experiments, we propose using a hold-out dataset $\mathcal{D}_{v}$ to achieve these goals, as represented by $M(f{_\theta}; \mathcal{D}_{v})$ in Equation~\eqref{eq:over-a}. As discussed in Section~\ref{intro}, the typical platform objectives consist of three sub-objectives: user usage time, engagement, and retention. We now describe how these sub-objectives can be computed using the soft top-$k$ technique and then describe the process of balancing them, considering their varying levels of learning difficulty.

\subsubsection{Objective Representation} Regarding the aspects of platform objectives, user usage time refers to the duration that a user spends in apps, while user engagement primarily measures the level of active interaction between users and the system. User retention, on the other hand, represents the proportion of users who revisit the system, and it is a long-term metric that cannot be directly computed.
From the perspective of recommendation results of a model $f_{\theta}$ on the dataset $\mathcal{D}_{v}$, we can represent these three aspects using the following quantities associated with the top-$k$ recommendation list: the cumulative watch time of the list (denoted as $M_1(f_{\theta};\mathcal{D}_{v})$), the total count of positive explicit feedback of the list (denoted as $M_2(f_{\theta};\mathcal{D}_{v})$), and the diversity of the list (denoted as $M_{3}(f_{\theta};\mathcal{D}_{v})$). 

To compute these list-wise quantities for a recommender model, we need all user feedback on the recommended videos. As such, for each user, we just consider generating a top-$k$ list from the videos that the user has interacted with in $\mathcal{D}_{v}$ instead of all videos. However, the top-$k$ operation is not differentiable, bringing difficulties in passing the gradient to the labeling model. To overcome the challenge, we employ a soft top-$k$ operation to compute these list-wise quantities, forming the representation of platform objectives.   

\vspace{+3pt}
\noindent $\bullet$ \textbf{Soft top-$k$ based objective computation}. To make the top-$k$ operation differentiable, we employ a smoothed approximation for it, namely SOFT~\cite{xie2020differentiable}. Specifically, SOFT approximates the output of the top-$k$ operation as the solution of an Entropic Optimal Transport problem, and the gradient of the SOFT operator could be efficiently computed based on the optimality conditions of the problem. 
So the $M_1(f_{\theta};\mathcal{D}_{v})$, $M_2(f_{\theta};\mathcal{D}_{v})$, and $M_3(f_{\theta};\mathcal{D}_{v})$ computed with the SOFT operator would become differentiable.

Specifically, for a user $u \in \mathcal{U}$, let $\mathcal{D}_{v}^{u}$ denote samples belonging to $u$.
For each sample $(\bm{x}, \bm{y}) \in \mathcal{D}_{v}^{u}$, we further denote its video duration by $x_{d} \in \mathcal{R}^{+}$, the positive explicit feedback by $\bm{y}_{e}\in\{0,1\}^{N}$ (where $N$ denotes the number of types of the feedback), and the watch time feedback by $y_{w}\in \mathcal{R}^{+}$. To compute the above list-wise quantities for a recommender model $f_{\theta}$, by inputting all predictions of samples in $\mathcal{D}_{v}^{u}$, 
 SOFT could generate an output for each sample to indicate whether it belongs to the top-$k$ samples ranking by the predictions. Formally, for a sample $(\bm{x},\bm{y})$, we denote the output as:

\begin{equation}\label{eq:soft}
\begin{split}
    \alpha_{u,\bm{x}} & = SOFT(\bm{x}; \{f_{\theta}(\bm{x}^{\prime})|(\bm{x}^{\prime},\bm{y})\in \mathcal{D}_{v}^{u} \} ) \\ 
    & = \begin{cases}
       1, \quad \text{if $f_{\theta}(\bm{x})$ is in the top-$k$ highest predictions,}  \\
       0, \quad \text{else,} 
    \end{cases}
\end{split}
\end{equation}
where $\{f_{\theta}(\bm{x}^{\prime})|(\bm{x}^{\prime},\bm{y})\in \mathcal{D}_{v}^{u}\}$ denotes all predictions for $\mathcal{D}_{v}^{u}$. 
Then $M_1(f_{\theta};\mathcal{D}_{v})$, $M_2(f_{\theta};\mathcal{D}_{v})$, and $M_3(f_{\theta};\mathcal{D}_{v})$ can be formulated as:
\begin{equation}\label{eq:sub-metrics}
\begin{split}
   M_1(f_{\theta};\mathcal{D}_{v}) = & \frac{1}{|\mathcal{U}|}\sum_{u\in\mathcal{U}} \sum_{(\bm{x},\bm{y})\in\mathcal{D}_{v}^{u}} \frac{\alpha_{u,\bm{x}}}{k} scale(y_{w}),  \\
   M_2(f_{\theta};\mathcal{D}_{v}) = & \frac{1}{|\mathcal{U}|}\sum_{u\in\mathcal{U}} \sum_{(\bm{x},\bm{y})\in\mathcal{D}_{v}^{u}} \frac{\alpha_{u,\bm{x}}}{k} \delta(sum(\bm{y}_{e})), \\
   M_3(f_{\theta};\mathcal{D}_{v}) = & \frac{1}{|\mathcal{U}|}\sum_{u\in\mathcal{U}} 
  \left( \sum_{(\bm{x},\bm{y})\in\mathcal{D}_{v}^{u}} \frac{\alpha_{u,\bm{x}}}{k}  (scale(x_{d}) - E_{w,k})^2 \right)^{-1/2},  
\end{split}
\end{equation}
where $E_{w,k} = \frac{1}{k} \sum_{(\bm{x},\bm{y})\in\mathcal{D}_{v}^{u}} \alpha_{u, \bm{x}} scale(x_{d})$ denotes the average of $x_{d}$ over top-$k$ samples, $|\mathcal{U}|$ denotes the number of unique users in $\mathcal{D}_{v}$, $scale(\cdot)$ is a scaling function; $\delta(\cdot)$ denotes the indicator function, 
and $\delta(sum(\bm{y}_{e}))=0$ if $\bm{y}_{e}$ is fully zero else $1$. 
Here, we use the standard deviation of the video duration to represent the diversity of the video duration. For simplicity, we omit the diversity regarding other features, which can be similarly computed if needed.

\textit{Overall objectives.} Then we merge $M_1(f_{\theta};\mathcal{D}_{v})$, $M_2(f_{\theta};\mathcal{D}_{v})$, and $M_3(f_{\theta};\mathcal{D}_{v})$ to form the overall platform objectives $M(f_{\theta};\mathcal{D}_{v})$ with each one as a sub-objective that should be maximized. Formally,
\begin{equation}\label{eq:merge-metrics}
 M(f_{\theta};\mathcal{D}_{v}) = M_1(f_{\theta};\mathcal{D}_{v}) + M_2(f_{\theta};\mathcal{D}_{v}) + M_3(f_{\theta};\mathcal{D}_{v}),
\end{equation}
for which we can compute the gradient $\nabla_{\theta}M(f_{\theta};\mathcal{D}_{v})$ along the chain: $M(f_{\theta};\mathcal{D}_{v})\rightarrow \alpha_{u,\bm{x}} \rightarrow \theta$ with the SOFT operator.

\subsubsection{Objective Balancing}
Direct merge to form $M(f_{\theta};\mathcal{D}_{v})$ would lead to imbalance problem. There are two reasons: 1) the scales of the three sub-objectives are different, making the large-scale sub-objectives dominate the labeling model learning; 2) the three sub-objectives may have different learning difficulties. Targeting the first reason, we have introduced a scaling scheme denoted by $scale(\cdot)$ when introducing the three sub-objectives in Equation~\eqref{eq:sub-metrics}. We next give the details of $scale(\cdot)$ and present the strategy to cope with the second reason.

\vspace{+5pt}
\noindent $\bullet$ \textbf{Scaling scheme.} 
To achieve scaling the three sub-objectives, we try to adjust the scale of watch time and video duration to $[0,1]$, as shown by $scale(y_{w})$ and $scale(x_{d})$ in Equation~\eqref{eq:sub-metrics}, to align with the scale of the explicit feedback. Then, as the sub-objectives are computed via the average or standard deviation of watch time, explicit feedback, or duration, we could make sure that the obtained sub-objectives scaled in $[0,1]$. Specifically, we take a modified min-max normalization, \ie $scale(\cdot)$. Considering the long-tail distribution of watch time and video duration, we consider using different granularity to normalize the head and tail values of the watch time or video duration. Taking watch time as an example, we compute $scale(y_{w})$ as follows 
(similarly for $scale(x_{d})$):
\begin{equation} \label{eq:scale}
    scale(y_{w})=
    \begin{cases}
        \frac{y_{w}}{w_{\beta}} \beta^\prime, \quad \text{if} \ 0 \leq y_{w} \leq w_{\beta}, \\
        1 - \left(1- \beta^\prime \right)  \frac{w_{max}-y_w}{w_{max}-w_{\beta}}, \quad \text{else},
    \end{cases}
\end{equation}
where $w_{max}$ denotes the max value of the watch time in $\mathcal{D}_{v}$, and $w_{\beta}$ denotes the $\beta$-th percentile of watch time distribution for $\mathcal{D}_{v}$, and $\beta^\prime = max( \frac{w_{\beta}}{w_{max}}, 1-\beta\%)$. 
This method can ensure that the head (small) values of watch time, which occupies $\beta$\% of the historical data, will not collapse to a too small interval after scaling, \ie making the interval greater than $[0,1-\beta\%]$. For example, we set $\beta\%=80\%$ in this work, then $scale(\cdot)$ would make sure that the watch time in $[0,w_{\beta}]$ scales to an interval greater than $[0,0.2]$.

\vspace{+5pt}
\noindent $\bullet$ \textbf{Dynamic balancing.} Despite having adjusted the scaling intervals uniformly across all sub-objectives, the direct amalgamation of them could potentially result in a state of non-equilibrium. This is attributed to the varying degrees of learning complexities they encounter and the potential interdependencies that might exist among them. To overcome the issues, we propose to reformulate the Equation~\eqref{eq:merge-metrics}  by dynamically merging the sub-objectives during the learning process as follows:
\begin{equation} \label{eq:dy-obj}
M(f_{\theta};\mathcal{D}_{v}) = \sum_{i=1}^{3} softmax(-\tau M_{i}(f_{\theta};\mathcal{D}_{v})) \cdot M_{i}(f_{\theta};\mathcal{D}_{v}),
\end{equation}
where $softmax(-\tau M_{i}(f_{\theta};\mathcal{D}_{v}))$ serves as a weight to control the contribution of $M_{i}(f_{\theta};\mathcal{D}_{v})$, and it would not involve computing gradients during learning. Formally, 
$$softmax(-\tau M_{i}(f_{\theta};\mathcal{D}_{v})) = \frac{\exp\left({-\tau M_{i}(f_{\theta};{\mathcal{D}_v})}\right)}{\sum_{j=1}^{3}\exp\left({-\tau M_{j}(f_{\theta};{\mathcal{D}_v})}\right) },$$
where $\tau$ is the temperature coefficient.  The smaller sub-objectives will get higher weights during learning, which could make our labeling model optimize all metrics simultaneously. 
The balanced objectives are the final platform objectives we use.
\section{Experiment}
We conduct experiments to answer four research questions:

\noindent\textbf{RQ1}: How does LabelCraft perform on real-world data compared to existing label generation methods?

\noindent \textbf{RQ2}: What is the impact of the individual components of LabelCraft on its effectiveness?

\noindent \textbf{RQ3}: Can LabelCraft effectively mitigate the duration bias?

\noindent \textbf{RQ4}: How do the specific hyper-parameters of LabelCraft influence its effectiveness?

\subsection{Experimental Settings}
\subsubsection{Datasets.}
We conduct extensive experiments on two real-world datasets: Kuaishou and Wechat. 
\begin{itemize}[leftmargin=*]
    \item \textbf{Kuaishou}. This private dataset is sourced from Kuaishou\footnote{\url{https://kuaishou.com/}}, a well-known short video-sharing platform in China. It consists of a collection of recommendation records within two weeks. Each data sample within the dataset contains feedback including watch time, likes, comments, and follows. Additionally, the dataset includes various features of users and videos, such as user viewing history, video duration, and video tags. To maintain the dataset's quality, we employ a 20-core filtering process, ensuring that each user/video has a minimum of 20 samples.

    \item \textbf{Wechat}. The dataset is released in the WeChat Big Data Challenge\footnote{\url{https://algo.weixin.qq.com/}}, which records user behaviors on short videos in two weeks. The dataset provides extensive user/video features and encompasses various types of interaction feedback, such as likes, follows, comments, and favorites. Considering that the dataset is denser, we use 120-core filtering to preprocess it.
\end{itemize}

The statistics of the preprocessed datasets are summarized in Table \ref{exp:data}. For each dataset, we split the data into three sets: training, validation, and testing. The first twelve days of data are allocated to the training set, while the last two days are used for the validation and testing sets, respectively. Additionally, to maintain consistency with Kuaishou, we only consider three types of positive explicit feedback for Wechat: comments, likes, and follows, which serve as the $\bm{y}_{e}$ in Equation~\eqref{eq:sub-metrics} for our method.

\begin{table}[]
\caption{Statistical details of the evaluation datasets.}
\vspace{-10pt}
\label{exp:data}
\begin{tabular}{ccccc}
\hline
Dataset  & \#User & \#Item & \#Intersection & Density \\ \hline
Kuaishou & 989    & 46,628 & 1,309,785      & 0.0284  \\
Wechat   & 15,075 & 12,992 & 4,778,294      & 0.0244  \\ \hline
\end{tabular}
\vspace{-10pt}
\end{table}

\subsubsection{Baselines.}
We compare the proposed LabelCraft method with the following label generation methods in recommendation:
\begin{itemize}[leftmargin=*]

    \item \textbf{WT}. It refers to directly using watch time as a label.

    \item \textbf{EF}. This method directly uses explicit feedback as the label. However, for a fair comparison to LabelCraft, we merge all explicit feedback as a label by computing $\delta(\sum(\bm{y}_{e}))$ in Equation~\eqref{eq:sub-metrics}. 

    \item \textbf{PC}~\cite{PC}. 
     In this method, the watch time of a video is compared to the video duration to determine whether a user has fully watched the video, forming the Play Completion label. 
    
    \item \textbf{PCR} \cite{PCR}. This method converts the watch time to the Play Completion Rate label, which represents the ratio of the user's watch time to the video's duration.

    \item \textbf{D2Q} \cite{D2Q}. 
    This is a SOTA labeling method considering duration debiasing. It generates quantiles-based labels from watch time, using backdoor adjustment for debiasing.

    \item \textbf{DVR} \cite{DVR}. 
    This is another SOTA label generation method with considering mitigating duration bias. It defines the Watch Time Gain label, which measures the relative user watch time on a short video compared with the average watch time of all users on short videos with similar duration.
    
\end{itemize}
Each method only uses its generated label to train a recommender model. However, our label would use all feedback to generate a label, and the optimized platform objectives include multiple \bym{sub-objectives}, which are related to both watch time and explicit feedback. Therefore, for a fair comparison, we further consider comparing the methods that fit multiple labels in a multi-task manner. Specifically, we mainly consider the following methods:
\begin{itemize}[leftmargin=*]
    \item \textbf{WT+EF}, which trains a recommender model with the labels generated by WT and EF using the Multi-gate Mixture-of-Experts (MMoE)~\cite{MMoE} framework. 
    \item \textbf{D2Q+EF}, which trains a recommender model with the labels generated by D2Q and EF using MMoE. 
    \item \textbf{DVR+EF}, which trains a recommender model with the labels generated by DVR and EF using MMoE.  
\end{itemize}
Here, with WT+EF as a reference, we mainly consider D2Q+EF and DVR+EF and ignore other combinations like PC+EF with two considerations: 1) D2Q and DVR are the SOTA label generation methods to deal with the watch time feedback, and 2) they are closer to our setting since D2Q and DVR consider pursuing both the cumulative watch time and debiasing (related to diversity in LabelCraft) and EF considers the positive explicit feedback.

\begin{table*}[]
\caption{
Performance comparison between the baselines and our LabelCraft, where the best results are highlighted in bold and sub-optimal results are underlined. “RI” indicates the relative improvement of LabelCraft over the corresponding baseline.
}
\vspace{-5pt}
\label{exp:main}
\begin{tabular}{cccccccccccccc}
\hline
\multirow{2}{*}{Method} & \multicolumn{5}{c}{Kuaishou}                                        &       &  & \multicolumn{6}{c}{Wechat}                                                 \\ \cline{2-7} \cline{9-14} 
                        & NWTG@10         & RI     & DS@10        & RI      & NEG@10          & RI    &  & NWTG@10         & RI     & DS@10       & RI      & NEG@10          & RI    \\ \hline
PC                      & 0.2121          & 41.6\% & 15           & 792.1\% & 0.7902          & 3.3\% &  & 0.4563          & 39.0\% & 10          & 134.4\% & 0.7776          & 7.2\% \\
PCR                     & 0.2493          & 20.5\% & 67           & 105.9\% & 0.8005          & 2.0\% &  & 0.4125          & 53.8\% & 12          & 100.3\% & 0.8109          & 2.8\% \\
WT                      & \underline{0.2939}          & 2.2\%  & 113          & 21.6\%  & 0.7991          & 2.2\% &  & 0.4972          & 27.6\% & 15          & 58.7\%  & 0.8201          & 1.7\% \\
D2Q                     & 0.2722          & 10.4\% & 122          & 12.4\%  & 0.7949          & 2.7\% &  & \underline{0.6202}          & 2.3\%  & \underline{23}          & 6.8\%   & 0.8191          & 1.8\% \\
DVR                     & 0.2814          & 6.7\%  & \underline{135}          & 1.8\%   & 0.7866          & 3.8\% &  & 0.5300          & 19.7\% & 18          & 32.2\%  & 0.8219          & 1.4\% \\
EF                      & 0.2557          & 17.5\% & 113          & 21.2\%  & \underline{0.8097}          & 0.8\% &  & 0.4593          & 38.1\% & 15          & 65.6\%  & \underline{0.8261}          & 0.9\% \\
WT+EF                   & 0.2631          & 14.2\% & 119          & 15.5\%  & 0.8000          & 2.1\% &  & 0.5195          & 22.1\% & 21          & 17.4\%  & 0.8205          & 1.6\% \\
D2Q+EF                  & 0.2800          & 7.3\%  & 111          & 23.4\%  & 0.7896          & 3.4\% &  & 0.5790          & 9.5\%  & 22          & 12.8\%  & 0.8197          & 1.7\% \\
DVR+EF                  & 0.2876          & 4.4\%  & 124          & 10.9\%  & 0.7862          & 3.9\% &  & 0.5698          & 11.3\% & 22          & 11.0\%  & 0.8232          & 1.3\% \\ \hline
LabelCraft              & \textbf{0.3003} & -      & \textbf{137} & -       & \textbf{0.8165} & -     &  & \textbf{0.6343} & -      & \textbf{24} & -       & \textbf{0.8338} & -     \\ \hline
\end{tabular}
\end{table*}

\subsubsection{Evaluation Metrics.}
To comprehensively evaluate the compared methods, diverse metrics are essential to evaluate the model performance on user usage time, engagement, and retention. Specifically, for the retention objective, the maximization of diversity within the top-$k$ recommendation list is pursued, and thus, the evaluation employs the standard deviation as a direct measure. We mainly measure the duration diversity, which can reflect the debiasing level for duration bias combined with other metrics, and denote the metric as {Duration Std@$k$} (DS@$k$). For the remaining objectives, they aim to maximize the cumulative watch time and positive explicit feedback within the top-$k$ recommendation list. To measure them, we design two novel metrics inspired by the widely used NDCG@k metric~\cite{Trirank}: Normalized Watch Time Gain@k (NWTG@$k$) and  Normalized Explicit-feedback Gain@k (NEG@$k$). For a recommendation list generated for user by the evaluated model, let $[y_w^1, \dots, y_w^k]$ and $[\bm{y}_e^1, \dots, \bm{y}_e^k]$ denote the corresponding lists of watch time and explicit feedback, where $y_{w}^{k}$ denotes the watch time of the $k$-th recommended video, we have:
\begin{equation}
\begin{split}
    &\text{NWTG}@k = \frac{\text{WTG}@k}{\text{WTG}^{\prime}@k}, \quad  \text{WTG}@k = \sum_{i=1}^{k} \frac{y_w^i}{\log_2(i+1)}, \\
    &\text{NEG}@k =  \frac{\text{EG}@k}{\text{EG}^{\prime}@k}  , \quad \text{EG}@k = \sum_{i=1}^{k} \frac{\delta(sum(\bm{y}_e^i))}{\log_2(i+1)},
\end{split}
\end{equation}
where $\text{WTG}^{\prime}@k$ denotes the the ideal $\text{WTG}@k$, which is the $\text{WTG}@k$ for the recommendation list directly generated according to $y_{w}$, similarly for $\text{EG}^{\prime}@k$, and $\delta(sum(\bm{y}_e^i)$ has the same means to that in Equation~\eqref{eq:sub-metrics}. Obviously, NWTG@$k$ and NEG@$k$ measure the total watch time and total count of positive explicit feedback with paying higher attention to top positions. Higher values for all metrics indicate better results, and the reported results have been averaged across users.

\subsubsection{Implementation Details.} 
To ensure fair comparisons, we utilize the DIN model as the backbone recommender model for all the methods. The implementation of the DIN architecture (\eg embedding and layer sizes) follows the specifications outlined in the DIN paper~\cite{DIN}. For the MMoE-based methods, we set the number of experts to 3 and the number of gates to 2. We implement each expert using an MLP (Multi-Layer Perceptron) module, searching the layer sizes in [$48 \times 48 \times 48 \times 1$, $48 \times 192 \times 48 \times 1$] and implement the gate model with a linear layer. Regarding our LabelCraft, we implement its labeling model $g_{\phi}$ as an MLP module with layer sizes of $48 \times 256 \times 256 \times 1$, which generates continuous labels from 0 to 1.

We optimize all models using the SGD optimizer~\cite{xiao2017attentional}, setting the maximum of optimization epochs to 1000. We employ the Binary Cross Entropy (BCE) loss \cite{NCF} for fitting binary labels and Mean Squared Error (MSE) loss \cite{ding2022interpolative} for fitting non-binary labels. The best models are determined based on the 
validation results,
using an early stopping strategy with the patience of 5. We leverage the grid search to find the best hyper-parameters. For our method and all baselines, we search the learning rate in the range of $\{$1$e$-1, 1$e$-2, 1$e$-3, 1$e$-4$\}$, the size of mini-batch in the range of $\{$4096, 8192, 16384$\}$, and the weight decay in $\{$1$e$-2, 1$e$-4, 1$e$-6$, 0\}$. 
For multi-task methods, the weight ratio to learning different labels is searched in [1:1, 1:2, 2:1]. For the special hyper-parameters of baselines, we search most of them in the ranges provided by their papers. Regarding our method, we set the hyper-parameter $\beta$ (in Equation \eqref{eq:scale}) to 80, the hyper-parameter $k$ of top-$k$ list metric (in Equation \eqref{eq:sub-metrics}) to 10, and search for the hyper-parameter $\tau$ of our method (in Equation \eqref{eq:dy-obj}) in the range of $\{$0, 0.1, \dots, 1.0$\}$. Besides, we directly leverage the validation set to serve as the hold-out dataset $\mathcal{D}_{v}$ for our LabelCraft.

\vspace{-10pt}
\subsection{Performance Comparison (RQ1)}
We first evaluate the overall performance of all compared methods on three aspects of the considered platform objectives. We summarize the results in Table~\ref{exp:main}, where we draw following observations:

\begin{itemize}[leftmargin=*]
    \item 
    LabelCraft consistently exhibits superior performance compared to the baselines across all evaluated aspects. This consistent superiority emphasizes the remarkable alignment between the labels generated by LabelCraft and the multi-aspect platform objectives. These results validate two critical points: 1) the great superiority of explicitly optimizing platform objectives during the label generation process, and 2) the effectiveness of LabelCraft in formulating and addressing the optimization problem.
    
    \item 
    PC and PCR, which manually generate new semantic labels based on watch time, consistently perform poorly compared to other methods in most cases. This indicates a misalignment between these methods and the platform's objectives. This is because the manual label generation rules used in PC and PCR tend to favor shorter videos~\cite{DCR}, leading to biased recommendations that deviate from the platform's intended objectives.
    
    \item 
    D2Q and DVR stand out among the baselines in terms of DS@10, indicating their ability to avoid bias towards videos with specific durations. However, their performance on NEG@10 is unsatisfactory. This can be attributed to their primary focus on addressing duration bias during label generation while overlooking explicit engagement feedback, such as "comments". On the other hand, EF prioritizes leveraging explicit feedback and achieves the best performance in terms of NEG@10 among the baselines. However, it performs poorly in other metrics. These results suggest the importance of taking into account comprehensive influencing factors to generate better labels.

    \item 
    Notably, the multi-task learning methods WT+EF, D2Q+EF, and DVR+EF demonstrate the ability to enhance specific metrics, but they inadvertently hinder other metrics when compared to optimizing individual labels. These findings confirm that even when amalgamating diverse (manually generated) labels through multi-task learning, aligning with platform objectives remains challenging. Thus, our approach LabelCraft stands out in its adaptive alignment with various aspects of the platform objectives, showcasing its superiority.
\end{itemize}

\subsection{Ablation Studies (RQ2)}
In order to optimize label generation efficiency in LabelCraft, we have incorporated the following essential design components:
1) designing three \bym{sub-objectives} to guide the learning process of our labeling model,
2) implementing a trainable labeling model to model diverse influencing factors, and
3) introducing a metric dynamic balancing strategy to ensure balanced learning across the \bym{sub-objectives}.
To validate the reasoning behind these design considerations, we thoroughly examine the impact of each crucial design element by comparing LabelCraft with its variants that disable these designs. Specifically, we introduce the following variants of LabelCraft:
\begin{itemize}[leftmargin=*]
    \item[-] \textbf{w/o S}, disabling the scaling scheme in our balancing strategy.
    \item[-]  \textbf{w/o B}, disabling the dynamic balancing in our balancing strategy.
    \item[-] \textbf{w/o WI}, removing watch time from our labeling model's inputs.
    \item[-] \textbf{w/o DI}, removing duration from our labeling model's inputs.
    \item[-] \textbf{w/o EI}, removing explicit feedback (`likes', `comments', `follows') from our labeling model's inputs.
    \item[-] \textbf{w/o WO}, removing the \bym{sub-objective} of the watch time, \ie $M_{1}(f_{\theta}; \mathcal{D}_{v})$ in Equation~\eqref{eq:sub-metrics}, from the optimization objective.
    \item[-] \textbf{w/o DO}, removing the \bym{sub-objective} of the diversity, \ie $M_{2}(f_{\theta}; \mathcal{D}_{v})$ in Equation~\eqref{eq:sub-metrics}, from the optimization objective.
    \item[-] \textbf{w/o EO}, removing the \bym{sub-objective} of the explicit feedback, \ie $M_{3}(f_{\theta};\mathcal{D}_{v})$ in Equation~\eqref{eq:sub-metrics}, from the optimization objective.
\end{itemize}
Table \ref{exp:abl} shows the comparison results on Kuaishou, from which we draw the following observations:

\begin{table}[t]
\caption{
Results of the ablation study for our LabelCraft method on Kuaishou.
}
\vspace{-5pt}
\label{exp:abl}
\begin{tabular}{cccc}
\hline
Method & NWTG@10 & DS@10 & NEG@10  \\ \hline
LabelCraft      & 0.3003  & 137  & 0.8165 \\
LabelCraft w/o B    & 0.2710 & 122 & 0.7960 \\
LabelCraft w/o S    & 0.2627  & 101  & 0.8165 \\
LabelCraft w/o WI   & {0.2677}  & 103  & 0.8151 \\
LabelCraft w/o DI   & 0.2935  & {127}  & 0.7919 \\
LabelCraft w/o EI   & 0.3237  & 124  & {0.7927} \\
LabelCraft w/o WO   & {0.2785}  & 112  & 0.8001 \\
LabelCraft w/o DO   & 0.3290  & {128}  & 0.7982 \\
LabelCraft w/o EO   & 0.3109  & 131  & {0.7962} \\ \hline
\end{tabular}
\vspace{-12pt}
\end{table}

\begin{itemize}[leftmargin=*]
    \item When LabelCraft disables the scaling scheme (w/o S) or dynamic balancing (w/o B), there is a decrease in performance across most metrics. These results confirm the importance of balancing and the significance of all balancing designs. Specifically, disabling the scaling scheme does not sacrifice the performance on explicit feedback metric NEG@10 but significantly sacrifices the others. This is because without scaling, the explicit feedback \bym{sub-objective} has a much smaller scale compared to the others, resulting in dynamic balancing assigning a higher weight to it and dominating the learning process. This suggests that scaling is the foundation for subsequent dynamic balancing.
    
    \item 
    Removing any \bym{sub-objective} from the optimization (w/o WO, w/o DO and w/o EO) would result in a performance decrease in at least one evaluation metric, particularly the evaluation metric corresponding to the removed \bym{sub-objective}. This confirms that optimizing each \bym{sub-objective} contributes to the achievement of the overall platform objective. However, removing a specific aspect of the objective may potentially improve the results on the metrics related to other aspects. This indicates the presence of a trade-off among different \bym{sub-objectives}, highlighting the need for balancing them accordingly.
    
    \item 
    Removing any input from the labeling model (w/o WI, w/o DI, and w/o EI) would lead to performance decreases on certain evaluation metrics. Comparing w/o WI and w/o WO, it's interesting that w/o WI has a more pronounced decrease on the corresponding watch time evaluation metric, NWTG@10. This pattern holds for w/o EI and w/o EO as well. These results highlight necessity of considering all influencing factors to generate labels.
    
\end{itemize}

\subsection{In-depth Studies (RQ3 \& RQ4)}

\subsubsection{ Debiasing Performance} 
\begin{figure}[t]
    \centering
    \subfigure[PC]{\includegraphics[width=0.23\textwidth,height=2.5cm]{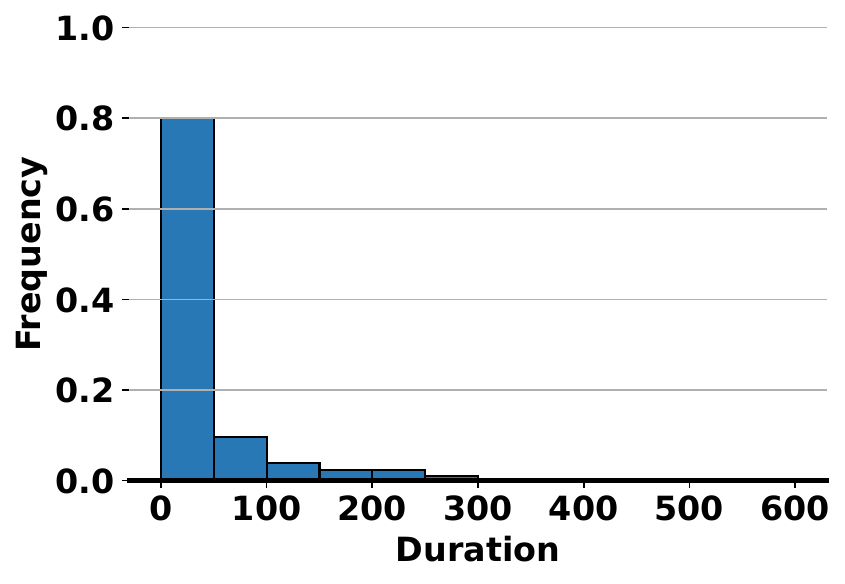}}
    \subfigure[D2Q]{\includegraphics[width=0.23\textwidth, height=2.5cm]{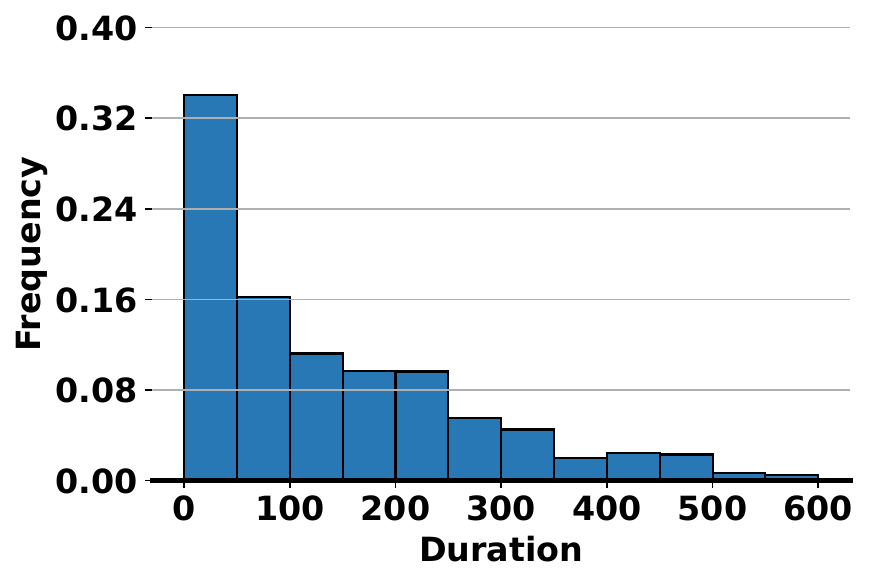}}
    \quad
    \subfigure[DVR]{\includegraphics[width=0.23\textwidth,height=2.5cm]{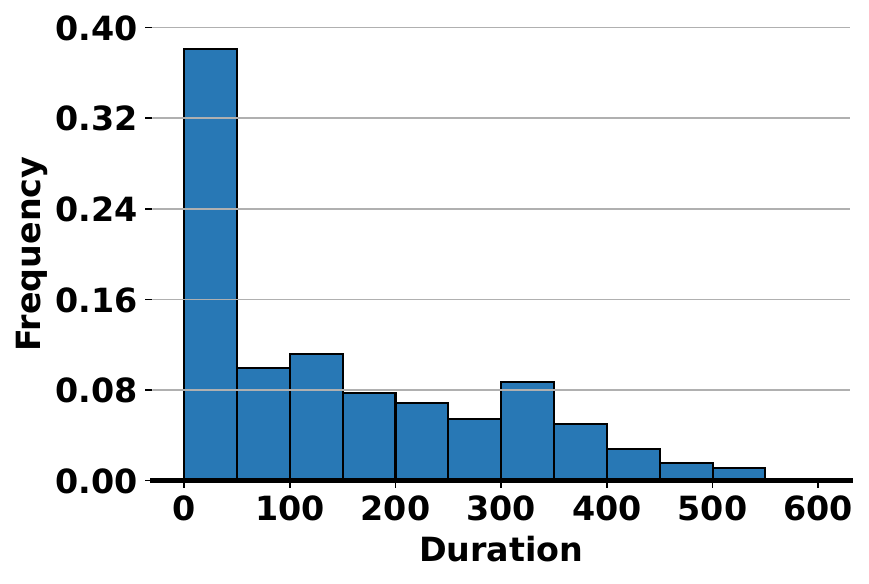}}
    \subfigure[LabelCraft]{\includegraphics[width=0.23\textwidth, height=2.5cm]{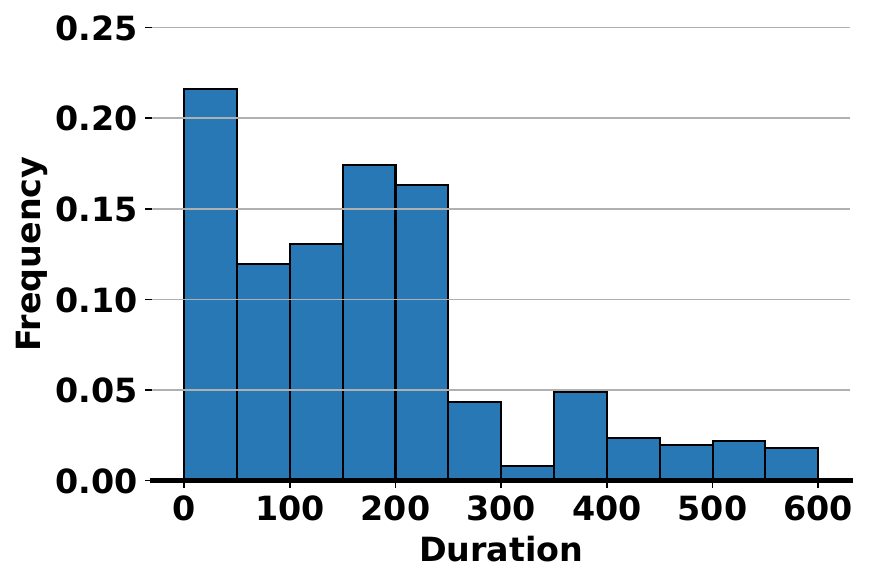}}
     \vspace{-10pt}
    \caption{
    Distribution of video duration in the top-$k$ recommended video lists generated by the PC, D2Q, DVR, and our LabelCraft for Kuaishou.
    }
    \label{fig:ana_duration}
    \vspace{-10pt}
\end{figure}

To assess the effectiveness of our method in mitigating duration bias, which is a notable strength shared by D2Q and DVR, we conduct an analysis experiment on Kuaishou following the approach outlined in the DVR paper~\cite{DVR}. Specifically, we analyze the histogram depicting the distribution of video duration in the top-$k$ recommended video lists generated by different models, and compare the results of the non-debiasing method PC, the debiasing methods D2Q and DVR, and our LabelCraft. The comparative results are summarized in Figure~\ref{fig:ana_duration}. 

According to Figure \ref{fig:ana_duration}, the PC method could not mitigate duration bias, 
as evidenced by its extremely skewed recommendation distribution and poor recommendation accuracy (see NWTG@10 and NEG@10 in Table~\ref{exp:main}). The recommendations of method DVR and D2Q become more balanced by debiasing as expected. Interestingly, our LabelCraft method demonstrates more balanced recommendations compared to DVR and D2Q. Additionally, it achieves better performance on accuracy-related metrics, as evidenced by the results on NWTG@10 and NEG@10 in Table~\ref{exp:main}. These findings indicate that LabelCraft effectively mitigates duration bias. The debiasing capability can be attributed to the alignment between the labels generated by LabelCraft and the platform objectives. When the platform objective is appropriately designed to be free of biases, it could guide the labeling model to generate bias-free labels.

\subsubsection{The effect of Hyper-parameter $\tau$}
 
\begin{figure}
    \centering
    \subfigure[NWTG@10 and NEG@10]{\includegraphics[width=0.25\textwidth]{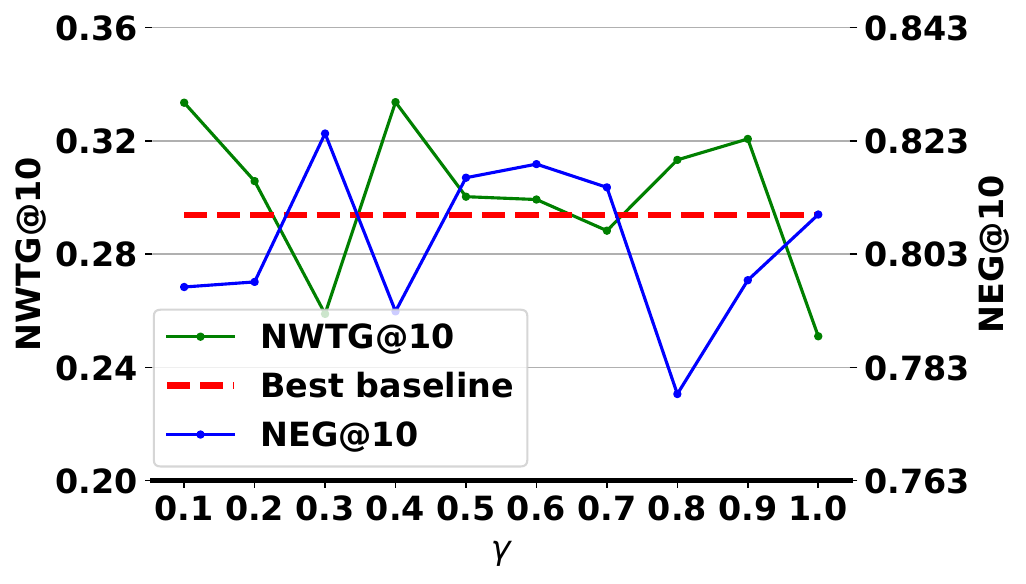}}
    \subfigure[DS@10]{\includegraphics[width=0.21\textwidth]{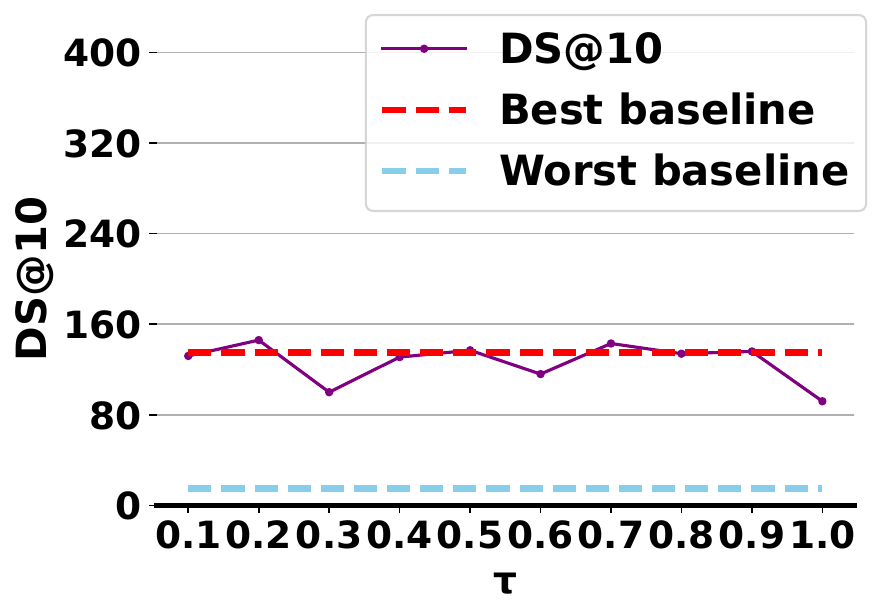}}
    \quad
    \vspace{-15pt}
    \caption{
    Performance of LabelCraft across different values of $\tau$, compared to the best results achieved by baselines.
    }
    \label{hyper_gamma}
    \vspace{-10pt}
\end{figure}

In our dynamic balancing design described in Equation~\eqref{eq:dy-obj}, the temperature coefficient ($\tau$) plays a crucial role in determining the strength of our dynamic balancing strategy. 
We conduct a study to investigate the impact of varying $\tau$ on LabelCraft's performance. Figure~\ref{hyper_gamma} illustrates the performance of LabelCraft for each metric, along with the best performance achieved by baselines on each metric, as $\tau$ is varied within the range of $[0,1]$ with a step size of 0.1. 
From Figure \ref{hyper_gamma}, it is evident that when $\tau=0.5$, LabelCraft consistently outperforms all baselines across all evaluation metrics. In the range of $\tau \in [0.5,0.7]$, the performance of LabelCraft remains relatively acceptable, either surpassing or comparable to the best baseline results on each metric\footnote{Within this range, LabelCraft achieves lower results on DS@10 compared to the best baseline. However, the impact on the balance of recommendation results remains small and the performance is close to the result of the second-best baseline.}. However, it is noted that consistent superiority over the best baseline performance cannot be maintained within this range. This suggests the importance of setting an appropriate value for $\tau$.

\section{Related Work}

\noindent $\bullet$ \textbf{Labeling in recommendation.}
Label generation is vital for developing an effective recommender model. 
The first area of research focuses on identifying varying levels of interest from the feedback (mainly watch time) and generating new semantic labels. For example, labels such as Play Completion (PC) \cite{PC}, Play Completion Rate (PCR) \cite{PCR}, and Effective View (EV) \cite{DML} has been designed to indicate the extent to which a user has engaged with a video. These labels play a crucial role in enhancing recommendations by providing a more nuanced understanding of users' interests. 
The second line of label generation work focuses on generating unbiased labels that can improve the accuracy of watch time prediction. 
DVR \cite{DVR} and D2Q \cite{D2Q} group and normalize the watch time according to the video duration to generate new labels. 
DML~\cite{DML} further considers the distribution of watch time.
Besides, some works focus on denoising \cite{denoising} and addressing clickbait issues \cite{clickbait} during label generation.
In contrast to these manual labeling works, we focus on introducing explicit optimization towards platform objectives to achieve automated label generation.

\vspace{+5pt}
\noindent $\bullet$ \textbf{Debiasing in recommendation.}
Recommender systems are susceptible to various biases, including position bias \cite{collins2018study, joachims2017accurately}, selection bias \cite{marlin2012collaborative, steck2013evaluation}, popularity bias \cite{zhang2021causal, wei2021model}, and duration bias \cite{D2Q, DVR, DML}. To tackle these biases, three main research lines have emerged. The first line of research involves utilizing unbiased data to guide model learning \cite{chen2021autodebias, li2023balancing, wang2021combating}, although acquiring such data can be expensive. The second line of research focuses on mitigating biases from a causal perspective, which can be categorized into intervention \cite{he2022causal, wang2021deconfounded, zhang2021causal} and counterfactual methods \cite{yang2021top, PCR, wei2021model}. The third line of research is the reweighting method \cite{wang2019doubly, schnabel2016recommendations, saito2020unbiased}, which utilizes inverse propensity scores \cite{schnabel2016recommendations} to adjust the training distribution and minimize bias. However, estimating these weights can pose challenges \cite{DCR}. Unlike the debiasing approaches mentioned earlier, our focus lies in eliminating the duration bias of labels by making the generated label align with the platform objectives.

\section{Conclusion}

This study highlights the problem of automatically generating reliable labels from raw feedback in short video recommendations. We propose LabelCraft, an innovative framework for automated labeling that formulates the label generation process as an explicit optimization problem. By learning a labeling model aligned with the platform objectives, LabelCraft demonstrates promising results in generating better labels, as evidenced by comprehensive experiments conducted on real data from popular platforms like Kuaishou and WeChat. However, industrial scenarios often involve numerous and complex platform objectives. Hence, future work aims to develop labeling methods capable of simultaneously aligning with more complex objectives. Meanwhile, we plan to decouple the learning of the labeling model from the recommendation learning process to enhance the transferability of the generated labels, enabling their compatibility across diverse model types, encompassing traditional models and the emerging LLM-based models~\cite{TALLRec, collm}.

\begin{acks}
    This work is supported by the National Key Research and Development Program of China (2022YFB3104701), the National Natural Science Foundation of China (62272437), and the CCCD Key Lab of Ministry of Culture and Tourism.
\end{acks}

\section*{Ethical Considerations}
In this section, we conscientiously contemplate the conceivable negative societal repercussions of our innovative method and discuss strategies aimed at alleviating these potential impacts.
\begin{itemize}[leftmargin=*]
    \item \textbf{Fairness}: 
    Considering that our approach to automatic label generation prioritizes optimizing operational metrics, there is a possibility of unintentional bias seeping into the labeling process. This bias can stem from biased optimization target design, leading to biased labels, and ultimately resulting in unfair user experiences with recommender systems. To address this concern, we intend to use a verification mechanism to optimize the target design and actively correct it when fairness violations occur.

    \item \textbf{Privacy}: The process of automated label generation entails the analysis and processing of user feedback data, thereby giving rise to apprehensions about user privacy. While we ardently uphold stringent data protection protocols, we explicitly acknowledge the potential for the reidentification of individuals through their feedback submissions. To mitigate this imminent risk, we shall employ meticulous anonymization strategies, curtail the retention of personally identifiable information, and seek adept legal counsel to ensure unwavering compliance with prevalent data privacy regulations.
    
    \item \textbf{Safety}: While the primary objective of LabelCraft revolves around augmenting user experience and engagement, the possibility of unintended consequences compromising user safety looms. This is especially salient in cases where the generated labels inadvertently endorse harmful or inappropriate content, consequently subjecting users to distressing material. To ensure an optimal level of user safety, we are steadfast in our resolve to implement stringent content moderation mechanisms, institute user-friendly reporting systems, and integrate human oversight to validate the precision and propriety of the generated labels.

\end{itemize}

\bibliographystyle{ACM-Reference-Format}
\balance
\bibliography{9_reference}

\appendix

\end{document}